\documentclass[sn-mathphys,Numbered]{sn-jnl}

\usepackage{graphicx}%
\usepackage{multirow}%
\usepackage{amsmath,amssymb,amsfonts}%
\usepackage{amsthm}%
\usepackage{mathrsfs}%
\usepackage{physics}

\usepackage[title]{appendix}%
\usepackage{xcolor}%
\usepackage{textcomp}%
\usepackage{manyfoot}%
\usepackage{booktabs}%
\usepackage{algorithm}%
\usepackage{algorithmicx}%
\usepackage{algpseudocode}%
\usepackage{listings}%

\providecommand{\ii}{\text{i}}

\providecommand{\vc}{\vb*}

\renewcommand{\Re}{\real}

\begin{document}

\title[]{Chiral forces in longitudinally invariant dielectric photonic waveguides}


\author[1]{\fnm{Josep} \sur{Martínez-Romeu}}
\equalcont{These authors contributed equally to this work.}
\author[1]{\fnm{Iago} \sur{Diez}}
\equalcont{These authors contributed equally to this work.}

\author[2]{\fnm{Sebastian} \sur{Golat}}

\author[2]{\fnm{Francisco J.} \sur{Rodríguez-Fortuño}}

\author[1]{\fnm{Alejandro} \sur{Martínez}} \email{amartinez@ntc.upv.es}

\affil[1]{\orgdiv{Nanophotonics Technology Center}, \orgname{Universitat Politècnica de València}, \orgaddress{\street{Camino de Vera, s/n Building 8F}, \postcode{46022}, \city{Valencia}, \country{Spain}}}

\affil[2]{\orgdiv{Department of Physics}, \orgname{King's College London}, \orgaddress{\street{Strand}, \postcode{WC2R 2LS}, \city{London},  \country{United Kingdom}}}





\abstract{\textbf{} Optical forces can be chiral when they exhibit opposite signs for the two enantiomeric versions of a chiral molecule or particle. Such forces could be eventually used to separate enantiomers, which could find application in numerous disciplines. Here, we analyze numerically the optical chiral forces arising in the basic element of photonic integrated circuitry: a dielectric waveguide with rectangular cross-section. Such waveguides are inherently lossless thus generating chiral forces that are invariant in the longitudinal direction and therefore enable enantiomeric separation over long (cm-scale) distances. Assuming Brownian motion in a liquid environment, we calculate first the force strength and time span needed to perform the separation of chiral nanoparticles as a function of the radii. Then we analyze the chiral forces produced by the fundamental quasi-TE guided mode in a silicon nitride waveguide and show that it can lead to enantiomeric separation via the transverse spin at short wavelengths (405 nm). At longer wavelengths (1310 nm), the proper combination of degenerate quasi-TE and quasi-TM modes would result in a quasi-circularly polarized mode with intrinsic chirality (helicity), leading to chiral gradient forces that also enable the enantiomeric separation of smaller nanoparticles. We report particle tracking simulations where the optical force field produced by a quasi-TE and a quasi-circular mode proved to separate enantiomers under a time span of two seconds. Our results suggest the viability of enantiomeric separation using simple photonic integrated circuits, though different wavelength windows should be selected according to the nanoparticle size.}

\maketitle

\section{Introduction}
Separation of enantiomers from racemic mixtures is essential in fields such as chemistry and pharmaceutics since the required performance is only exhibited by an enantiomer with a certain handedness (or chirality). Usually, methods based on chemical interactions, such as chiral High-Performance Liquid Chromatography (HPLC), are employed to separate enantiomers starting from racemic mixtures. However, such chemical methods are usually slow, expensive, and molecule-dependent \cite{Jacques1981}. An interesting alternative would be the use of optical chiral forces \cite{Genet2022}: since the chiral force exerted by light changes its sign when acting on enantiomers of different handedness \cite{Zhao2017}, it could ultimately lead to light-driven enantiomeric separation overcoming many of the limitations of chemical methods. 

There have been many recent theoretical and simulation works addressing the separation of enantiomers using light \cite{Canaguier-Durand2013, Hayat2015,Canaguier-Durand2015,Zhang2017,Cao2019,Zheng2020,Zhang2021}. Indeed, there have been several experiments demonstrating optically-induced separation, though for relatively large nanoparticles and nanostructures \cite{Tkachenko2014,Magallanes2018,Shi2020}. In all cases, free-space propagating beams are used, either being reflected at dielectric interfaces \cite{Hayat2015} or interfering with other beams \cite{Zhang2017} to produce the transverse optical spin that generates the required forces. Alternatively, one may think of using guided light for separation purposes, taking advantage of the enhancement of chiral interaction because the light is confined in subwavelength cross-sections over relatively long distances (ideally infinite for a lossless waveguide). One possibility is the use of optical nanofibers with cylindrical cross-section, as recently noticed by Golat \textit{et al.} \cite{Golat2023}. Another possibility would be the use of dielectric waveguides that can be created by lithography in photonic integrated circuits (PICs) and can exhibit either transverse \cite{Espinosa-Soria2016} or longitudinal spin \cite{Vázquez‐Lozano2020} for guided modes. Remarkably, such waveguides can be massively integrated into PICs and, in the case of silicon-related materials, fabricated in large volumes using low-cost processes. Recently, several approaches to separating enantiomers using integrated waveguides have been presented \cite{Fang2021,Liu2023,Fang2023}. However, in all of them, the separating chiral forces are not kept over long propagation distances, just missing this clear advantage of PICs over free-space approaches.

In this work, we analyze the chiral separation properties of the most simple photonic integrated structure: a lossless dielectric waveguide with rectangular cross-section on a lower-index substrate. We consider silicon nitride (SiN) as the material to build the waveguide core since it is transparent from telecom wavelengths down to the ultraviolet. The refractive index of SiN is large enough to ensure tight guiding when the core lies on a silicon dioxide substrate and is surrounded by water. Remarkably, waveguides with very low propagation loss ($<$ 1dB/cm) can be fabricated using mature tools and processes \cite{Xiang2022}. We first calculate the order of magnitude of the required chiral forces to perform the enantiomeric separation of particles under a reasonable time assuming Brownian diffusion of the target particles in a liquid environment. Then, we calculate numerically the electric and magnetic fields at different wavelengths from 405 to 1310 nm to obtain the optical forces using well-established equations \cite{Golat2023}. We show that at short wavelengths ($\approx{405}$ nm) lateral chiral forces arising from the transversal spin of the guided quasi-TE mode \cite{Espinosa-Soria2016} can overcome the achiral forces and be used for separation of chiral particles of 80 nm radius. At longer wavelengths ($\approx{1310}$ nm), this lateral force becomes much smaller but we combine the quasi-TE and quasi-TM modes of the waveguide with a proper $90^\circ$ phase shift between them to generate a quasi-circularly polarized (quasi-CP) mode \cite{Vázquez‐Lozano2020}. This mode produces a large transversal chiral gradient force that could separate nanoparticles of 52 nm radius and potentially molecules with a size of the order of 1 nm.  Our results suggest that long SiN waveguides are simple but realistic structures towards achieving enantiomeric separation within seconds using guided light in PICs.



\section{Optical forces exerted on small chiral particles}
The electromagnetic field of light carries momentum that can be transferred to a particle through the action of an optical force, and consequently, cause its motion. In this work, we restrict to study the motion of small chiral particles (whose size is smaller than the wavelength of light) subjected to optical forces. A small particle is fully characterized by its electric dipole moment, $\vc{p}$, which can be thought of as the separation of positive and negative charges, and its magnetic dipole moment, $\vc{m}$, which represents the overall current loop within the particle. The time-averaged force $\vc{F}$ that the electromagnetic field exerts on a small particle is: \cite{Chaumet2009,Nieto2010,Hayat2015,Golat2023}


\begin{equation}\label{eq:general_force}
    \!{\vc{F}}=\frac{1}{2}\Re\Big[\!\underbrace{\vphantom{\frac{k^3}{6\pi}}(\grad\!\otimes\!\vc{E})\vc{p}^\ast\!+\!\mu(\grad\!\otimes\!\vc{H})\vc{m}^\ast}_\text{interaction}\underbrace{-\,\frac{k^4\eta}{6\pi}(\vc{p}^\ast\!\!\cp\!\vc{m})}_\text{recoil}\!\Big],\!\!
\end{equation}


In this expression, $\vc{E}$ and $\vc{H}$ are respectively the electric and the magnetic field at the position of the particle, $\eta=\sqrt{\mu/\varepsilon}$ is the impedance of the surrounding medium, $\varepsilon$ is the electric permittivity of the medium, $\mu$ is the magnetic permeability of the medium, $k=2\pi/\lambda$ is the wavenumber, and $\lambda$ is the wavelength of light. The dipole moments of a particle arise due to its interaction with the electric and magnetic fields of light, and are obtained as follows:
%
\begin{equation}\label{eq:dipolar_moments}
\begin{split}
\vc{p} & = \alpha_e \varepsilon \vc{E} + \ii\frac{1}{c}\alpha_c \, \vc{H} \\
\vc{m} & = \alpha_m \vc{H} - \ii\frac{1}{\eta}\alpha_c \vc{E}
\end{split}
\end{equation}
\noindent where $(\alpha_e, \alpha_m, \alpha_c)$ are the dynamic electric, magnetic and chiral polarizabilities of the particle, and $c=1/\sqrt{\mu\varepsilon}$ is the speed of light in the medium. The static polarizabilities of a spherical particle of radius $r$ can be modelled using the generalized Clausius-Mossotti expressions \cite{Canaguier-Durand2015,Genet2022}:
\begin{equation}\label{eq:polarizabilities}
\begin{split}
    \alpha_{0e} & = 4\pi r^3 \frac{(\varepsilon_p-\varepsilon_m)(\mu_p+2\mu_m)-\kappa^2}{(\varepsilon_p+2\varepsilon_m)(\mu_p+2\mu_m)-\kappa^2} \\
    \alpha_{0m} & = 4\pi r^3 \frac{(\varepsilon_p+2\varepsilon_m)(\mu_p-\mu_m)-\kappa^2}{(\varepsilon_p+2\varepsilon_m)(\mu_p+2\mu_m)-\kappa^2} \\
    \alpha_{0c} & = 12\pi r^3 \frac{\kappa}{(\varepsilon_p+2\varepsilon_m)(\mu_p+2\mu_m)-\kappa^2}
\end{split}
\end{equation}

\noindent where $(\varepsilon_p, \mu_p,\kappa)$ refer to the relative permittivity, relative permeability, and chirality parameter of the particle, and $(\varepsilon_m, \mu_m)$ refer to the relative permittivity and permeability of the non-chiral background medium. A radiation damping has to be added to expressions~(\ref{eq:polarizabilities}) to satisfy the conservation of energy \cite{Sersic2011}. This so-called radiative correction is often applied incorrectly in the literature for chiral particles, because it is only applied to the electric and magnetic polarizabilities, thus, neglecting the correction for the chiral polarizability. However, as shown by 
Sersic \textit{et al.} \cite{Sersic2011}, the tensor radiative correction has to be applied to the full $6\times6$ square polarizability matrices, yielding the following expressions for the dynamic polarizabilities \cite{Golat2023}, which are used for calculating the dipole moments in Eq.~(\ref{eq:dipolar_moments}).
    %
    %
%
\begin{equation} \label{eq:radiative_correction}
\begin{split}
    \!\!\!\alpha_\text{e}&=\frac{\alpha_\text{0e}
    -\ii\frac{ k^3}{6 \pi}(\alpha_\text{0c}^2-\alpha_\text{0e}\alpha_\text{0m})}{1+\qty(\frac{k^3}{6 \pi})^2\qty(\alpha_\text{0c}^2-\alpha_\text{0e} \alpha_\text{0m})-\ii \frac{k^3}{6 \pi}\qty(\alpha_\text{0e}+\alpha_\text{0m})},\!\!\!\\
    \!\!\!\alpha_\text{m}&=\frac{\alpha_\text{0m}
    -\ii\frac{ k^3}{6 \pi}(\alpha_\text{0c}^2-\alpha_\text{0e}\alpha_\text{0m})}{1+\qty(\frac{k^3}{6 \pi})^2\qty(\alpha_\text{0c}^2-\alpha_\text{0e} \alpha_\text{0m})-\ii \frac{k^3}{6 \pi}\qty(\alpha_\text{0e}+\alpha_\text{0m})},\!\!\!\\
    \!\!\!\alpha_\text{c}&=\frac{\alpha_\text{0c}
    }{1+\qty(\frac{k^3}{6 \pi})^2\qty(\alpha_\text{0c}^2-\alpha_\text{0e} \alpha_\text{0m})-\ii \frac{k^3}{6 \pi}\qty(\alpha_\text{0e}+\alpha_\text{0m})}.\!\!\!
\end{split}
\end{equation}

With Eqs.~(\ref{eq:general_force},\ref{eq:dipolar_moments},\ref{eq:polarizabilities},\ref{eq:radiative_correction}) we are ready to compute the time-averaged optical force due to a monochromatic electromagnetic field on a small particle. However, we gain more insight by developing the expression of the force from Eq.~(\ref{eq:general_force}). The force can be split into several terms which depend on the following six time-averaged local field properties: electric energy density $W_e$, magnetic energy density $W_m$, helicity density $\mathfrak{G}$, electric spin density $\vc{S}_e$, magnetic spin density $\vc{S}_m$ and complex Poynting vector $\vc{\varPi}$:
\begin{equation}
\begin{split}
W_e & = \frac{1}{4}\varepsilon |\vc{E}|^2 \;\;  \left[\frac{J}{m^3}\right]\\
W_m & = \frac{1}{4}\mu |\vc{H}|^2 \;\;  \left[\frac{J}{m^3}\right]\\
\mathfrak{G} & = \frac{1}{2\omega c} \imaginary \left( \vc{E} \cdot \vc{H}^* \right) \;\;  \left[\frac{J\cdot s}{m^3}\right]\\
\vc{S}_e & = \frac{1}{4\omega} \imaginary \left( \varepsilon \ \vc{E}^*\, \times \vc{E} \right) \;\;  \left[\frac{J\cdot s}{m^3}\right]\\
\vc{S}_m & = \frac{1}{4\omega} \imaginary \left(\mu \, \vc{H}^*\times \vc{H} \right) \;\;  \left[\frac{J\cdot s}{m^3}\right]\\
\vc{\varPi} & = \frac{1}{2}\vc{E} \times \vc{H}^* \;\;  \left[\frac{W}{m^2}\right]\\
\end{split}
\end{equation}

The fully developed expression of the optical force acting on a particle is shown below, which has been split into the chiral and achiral terms, depending on whether the terms are a function of the chiral polarizability or not, respectively \cite{Golat2023}:
\begin{equation}\label{eq:all_forces}
\begin{split}
    \vc{F}_{\text{chiral}} = & \underbrace{\vphantom{\frac{1}{c}}\;\omega \; \real(\alpha_c) \nabla \mathfrak{G}}_{\text{helicity gradient}} - \underbrace{\frac{1}{c} \imaginary(\alpha_c)\nabla \times \real \vc{\varPi}}_{\text{vortex}} \\
    + & \underbrace{\left( \; 2k \imaginary(\alpha_c) - \frac{k^4}{3\pi} \real(\alpha_e^*\alpha_c) \right) \omega  \vc{S}_e}_{\text{electric spin}} \
    + \underbrace{\left( \; 2 k \imaginary(\alpha_c) - \frac{k^4}{3\pi} \real(\alpha_c^*\alpha_m) \right) \omega  \vc{S}_m}_{\text{magnetic spin}} \\
    \vc{F}_{\text{achiral}} = & \; \underbrace{\real(\alpha_e)\nabla W_e}_{\text{electric gradient}} + \underbrace{\real(\alpha_m)\nabla W_m}_{\text{magnetic gradient}} - \ \underbrace{\omega \nabla \times \left(\imaginary(\alpha_e)\vc{S}_e + \imaginary(\alpha_m)\vc{S}_m \right)}_{\text{spin-curl}} \\
    + & \underbrace{\left(\frac{k}{c}\imaginary(\alpha_e + \alpha_m) -\frac{k^4}{6\pi} \frac{1}{c}\left( \real(\alpha_e^*\alpha_m) + |\alpha_c|^2\right)\right) \real\vc{\varPi}}_{\text{radiation pressure}} \\
    - & \underbrace{\frac{k^4}{6\pi}\frac{1}{c}\imaginary(\alpha_e^*\alpha_m)\imaginary \vc{\varPi}}_{\text{flow}}
\end{split}
\end{equation}
%

%
All the forces exhibit an inherent dependency on the volume of the particle because the polarizabilities are proportional to $r^{3}$. In addition, some of the forces show explicit dependency on the wavelength (or wavenumber) too, being proportional to $r^3/\lambda$ or $r^6/\lambda^4$, such as the electric and magnetic spin forces and radiation pressure force, whereas others, such as the gradient forces, do not. Therefore, the dominance of chiral forces over achiral forces vary depending on particle size and wavelength. A more detailed analysis of the chiral and achiral optical forces exerted on small chiral particles is discussed by Golat \textit{et al.} \cite{Golat2023}.

\section{Optically driven motion of small chiral particles in a fluid}
To study under what circumstances the optical enantioseparation is possible, we first need to examine the motion of particles in a fluid under the influence of an external chiral optical force field.
%
%
To this end, we consider the following assumptions for this system: there is no net fluid flow, and the mass of the particles is negligible so that the viscous forces dominate over the inertial forces. Under these considerations, the motion of the particle is driven by a combination of the external optical force, $\vc{F}$, the friction or drag force due to the viscosity of the fluid (that is opposite to the movement of the particle), and the force arising from the stochastic collisions of the smaller fluid molecules with the particle (Brownian motion). The variation of the particle position, $\vc{x}(t)$, in time is governed by the overdamped Langevin equation \cite{Kravets2019,Schnoering2021}
\begin{equation}\label{eq:Langevin}
       \underbrace{\vphantom{} \gamma \frac{\text{d}\vc{x}}{\text{d}t}}_{\text{friction}} = \underbrace{\vphantom{\frac{d\vc{r}}{dt}} \vc{F}}_{\text{optical}} + \underbrace{\vphantom{\frac{d\vc{r}}{dt}} \gamma\;\sqrt{2D}\ \boldsymbol{\xi}(t)}_{\text{stochastic}}        
\end{equation}
\noindent where $D=k_B T/\gamma$ is the particle's diffusion coefficient within the bulk of the fluid, $k_B$ is the Boltzmann constant and $T$ is the absolute temperature. Under the assumption that the particles have spherical shape of radius $r$, the friction coefficient can be expressed as $\gamma = 6\pi \eta r$ (Stoke's law of friction), where $\eta$ is the dynamic viscosity of the fluid.
Eq.~(\ref{eq:Langevin}) is a stochastic differential equation that can be solved numerically with the Euler-Maruyama integration scheme to track the location of the particle upon an increment in time $\Delta t$ \cite{Kloeden1992,Kieninger2021}. The solution can be expressed as:
\begin{equation} \label{eq:maruyama}
\begin{split}
    x^{(m+1)} & =  x^{(m)} + \frac{F_x^{(m)}}{\gamma}\Delta t + \sqrt{2D_{x,\perp}^{(m)}\Delta t} \; N_x(0,1) \\
    y^{(m+1)} & = y^{(m)} + \frac{F_y^{(m)}}{\gamma}\Delta t + \sqrt{2D_{y,\perp}^{(m)}\Delta t} \; N_y(0,1) \\
    z^{(m+1)} & = z^{(m)} + \frac{F_z^{(m)}}{\gamma}\Delta t + \sqrt{2D_{z,||}^{(m)}\Delta t} \; N_z(0,1)
\end{split}
\end{equation}
where the upper index $(m)$ represents the $m-$th instant of time, and $N_{x/y/z}(0,1)$ represents three independent standard normal distributions with $0-$mean and $1-$variance. The force field $(F_x^{(m)},F_y^{(m)},F_z^{(m)})$ and the diffusion coefficients are evaluated at the location of the particle ($x^{(m)},y^{(m)},z^{(m)}$). The diffusion coefficients are modified with respect to their bulk value ($D$) as the particle moves near a boundary, such as the interface of the fluid with the waveguide or the walls of a microfluidic channel. This modification depends on whether the movement of the particle is perpendicular or parallel to the boundary \cite{Swan2007}. More details are given in Appendix~\ref{sec:particle_tracking}. Eq.~(\ref{eq:maruyama}) is used in Section~\ref{sec:results} to follow the trajectories of individual particles throughout a microfluidic channel subjected to the force field exerted by the waveguide mode.

\begin{figure}
    \centering
    \includegraphics[width=1\textwidth]{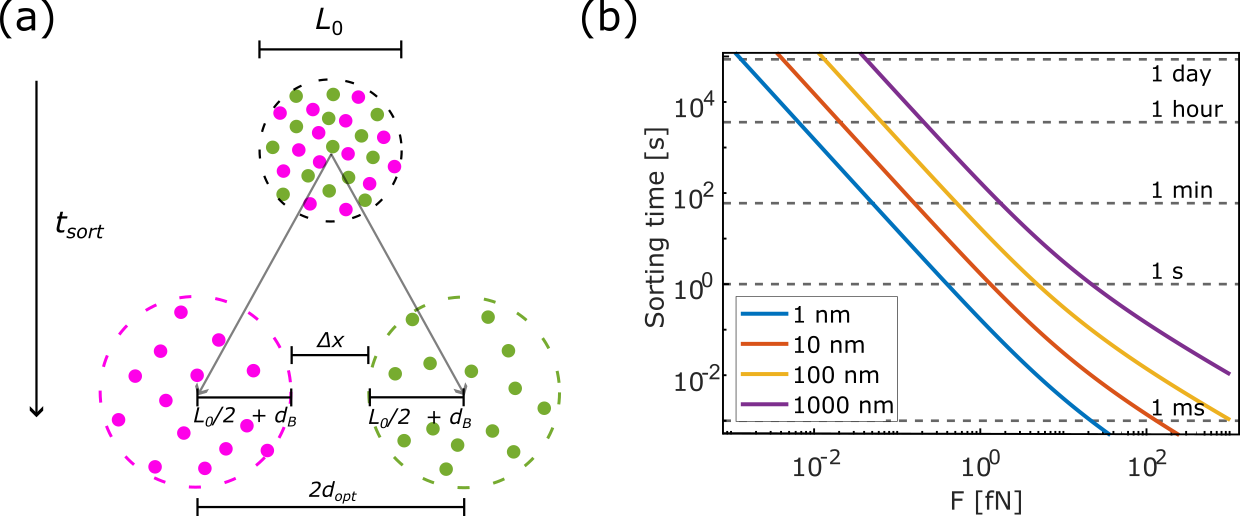}
    \caption{(a) Schematic showing the combined action of the translation of the cloud of enantiomers due to the optical force ($d_{\text{opt}}$) and the increase in the cloud size due to Brownian motion ($d_{\text{B}}$). After a sorting time ($t_{\text{sort}}$), the clouds are separated by a distance of $\Delta x$. Schematic adapted from \cite{Kravets2019}. (b) Sorting time of the enantiomer clouds as a function of the modulus of the separating optical force. Particles of different radii display different sorting time curves, from 1 nm to 1000 nm.}
    \label{fig:sorting}
\end{figure}

The magnitude of the chiral optical forces required for enantiomeric separation and the needed sorting time can be estimated upon further assumptions. For that we examine the movement of a cloud of particles under the influence of an external optical force field in the bulk, i.e. in an infinite system with no boundaries. We followed the derivation from Kravets \textit{et al.} \cite{Kravets2019}.
The enantiomers are modelled as non-interacting spherical particles and are assumed to be initially mixed and distributed within a spherical cloud of diameter $L_0$, as shown in Fig.~\ref{fig:sorting}(a). Let us assume the optical force field is uniform in space, constant in time, oriented along the $x-$direction, and dominantly chiral, i.e. the achiral part of the force is negligible compared to the chiral part ($F=|F_{\text{chiral}}|>>|F_{\text{achiral}}|$). The chiral optical force $F$ exerted on the particles moves each enantiomer in opposite directions, thus effectively separating the initial racemic mixture into two separate clouds of particles. Upon an interval of time $t$, each enantiomer cloud is displaced a distance $d_{\text{opt}}=Ft/\gamma$ due to the optical force. In addition to this displacement, the radius for each enantiomer cloud is expected to increase on average $d_\text{B}=\sqrt{2Dt}$ due to Brownian motion. These two simultaneous processes of motion are schematized in Fig.~\ref{fig:sorting}(a).
The condition to achieve a separation $\Delta x$ between the two clouds within a sorting time $t_{\text{sort}}$ gives rise to the following equation whose derivation is shown in detail in the Appendix~\ref{sec:sorting_time}.
\begin{equation}\label{eq:sorting_time}
t_{\text{sort}}= 3 \pi \eta k_\text{B}T\frac{r }{F^2}\left( 1 + \sqrt{1 + \frac{F(L_0+\Delta x)}{k_\text{B}T}} \right) ^2
\end{equation}

In the integrated waveguide system, the particles flow along a microfluidic channel and the force field is generated by a waveguide. Thus, the actual force field is not spatially uniform and there are boundaries that enclose the region of space where the particles can move and that modify the diffusion coefficient value. However, Eq.~(\ref{eq:sorting_time}) provides a quite accurate estimation of the sorting time for a longitudinally invariant dielectric waveguide-microchannel system. We assume the initial extension of the racemic mixture cloud is $L_0=1$ $\mu$m. This is achievable with the well-known microfluidics technique called hydrodynamic flow focusing \cite{Knight1998}, where two lateral flows can control the width of the middle fluid channel where the particles are suspended. The fluid is composed mainly of water at temperature $T$=293 K so the dynamic viscosity is $\eta=10^{-3}$ Pa $\cdot$ s \cite{Kravets2019}.

Fig.~\ref{fig:sorting}(b) shows the sorting time for particles of different radii, calculated with Eq.~(\ref{eq:sorting_time}) for $\Delta x=0$, which marks the starting point of separation. Stronger forces are needed to sort larger particles for the same value of sorting time. Moreover, larger particles take longer to be sorted for the same value of the force. We emphasize that this sorting time is obtained for a force field that is constant in time and uniform in space, which is not the actual situation of the force field generated by a waveguide system. In the latter case, the forces are stronger at distances closer to the waveguide and decay with the distance due to the evanescent field of the mode in the fluid. 
The usefulness of this graph is to know what range of optical force magnitude the waveguide system needs to generate to sort particles under a reasonable time: 1 ms $-$ few hours. We must therefore look for integrated waveguides that generate optical chiral forces within the range of $10^{-3}$ fN$ - 10^3$ fN for sorting particles of radii between 1 nm and 1000 nm. Once we have designed those waveguide systems, we use the particle tracking algorithm (Eq.~(\ref{eq:maruyama})) to test the actual enantiomeric separation that our waveguides can reach.

%
\section{Description of the integrated photonic waveguide}
We consider a photonic strip waveguide made with a SiN core (refractive index $n\approx2$) on top of a SiO$_2$ $(n=1.4468)$ substrate, and surrounded by water $(n=1.33)$ as the system that produces the force field responsible for the enantiomeric sorting; as shown in Fig.~\ref{fig:cross-section}(a). The use of SiN has several practical advantages such as transparency at visible and near-infrared wavelengths, a relatively large refractive index to ensure tight localization of the fields in the waveguide core, and its processing with a mature silicon technology to produce low-loss waveguides \cite{Xiang2022}. Moreover, SiN is particularly appropriate for applications requiring immersion in fluid, such as photonic biosensing \cite{Castello21}.

In order to obtain the electric and magnetic fields of the guided modes, the Maxwell's equations are solved in the waveguide using the finite element method implemented by the FemSIM solver in the commercial software RSoft (Synopsis). The software computes the eigenmode of the cross-section of the waveguide system, which is assumed to be invariant along the longitudinal direction (translational symmetry along the optical axis, i.e. $z-$axis). The resulting electric and magnetic fields are plugged into Eq.~(\ref{eq:all_forces}) to obtain the optical forces per amount of power guided by the mode. We assume 20 mW of power in our simulations which, despite being a high power level for integrated optics, can be attained using standard semiconductor continuous wave laser without causing material damage. 
Noticeably, the forces depend on the characteristics of the particle, which are modelled by the polarizabilities (Eqs.~(\ref{eq:polarizabilities}) and ~(\ref{eq:radiative_correction})). The particle's material is modelled with a relative permittivity of $\varepsilon_p=2$, relative permeability $\mu_p=1$, and chirality parameter $\kappa =\pm 0.5$. These values have been widely used to characterize chiral nanoparticles in the literature \cite{Hayat2015,Li2021}. The medium (water) is modelled with the values $\varepsilon_m=1.77$ and $\mu_m=1$. We take into account the following two conditions for the design of the waveguide towards enantiomeric separation: (I) $|F_{\text{chiral}}|>|F_{\text{achiral}}|$, to ensure that chiral forces dominate over achiral ones (although it is not a necessary condition for achieving separation); and (II) $|F_{\text{chiral}}|\sim  10^{-3}$ fN$ - 10^3$ fN, according to our calculations from Fig.~\ref{fig:sorting}.

After the total optical force field is computed, we use the particle tracking algorithm to track the position of an individual particle for each enantiomer for either 1 or 2 seconds throughout a hypothetical microfluidic channel surrounding the waveguide. Due to the stochastic nature of the Brownian motion, we repeat this tracking 500 times from the same initial position to do a statistical analysis of the enantioseparation process. The initial position for all particles is at $x=0$ and at a middle height between the top surface of the waveguide and the ceiling of the microchannel. From the final position of the particles, we calculate the enantiomer fraction (EF) for each enantiomeric cloud. More details about the particle tracking algorithm and statistical analysis can be found in Appendix~\ref{sec:particle_tracking}.

\section{Results}\label{sec:results}
\begin{figure}
    \centering
    \includegraphics[width=1\textwidth]{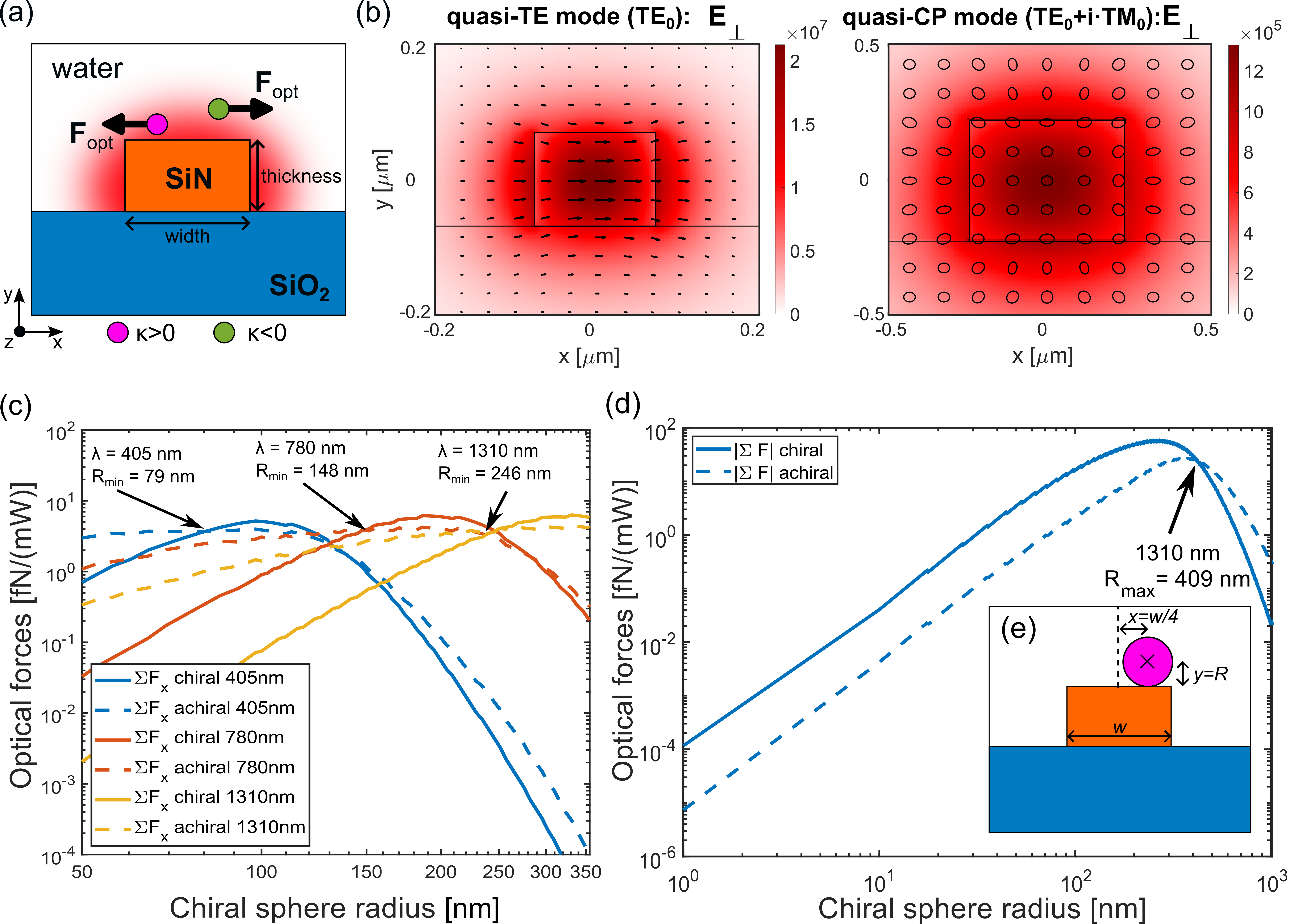}
    \caption{(a) Schematic of the waveguide cross-section representing the opposite action of the total optical force exerted by the guided mode (in red) onto particles with opposed chirality. (b) Transversal electric field intensity (in color map) and polarization (in arrow or ellipse map) of the quasi-TE mode and of the quasi-CP mode. The handedness of the polarization ellipses does not change its sign throughout the cross-section. (c) Net chiral and achiral force along the x-direction that a quasi-TE mode in a SiN strip waveguide exerts on a particle, depending on its size. This is calculated for three different SiN waveguides, each one operating at a different wavelength: 405 nm, 780 nm, or 1310 nm. The SiN cross-section size (width $\times$ thickness) of the waveguides are: 0.151 $\mu$m $\times$ 0.139 $\mu$m (for $\lambda=405$ nm), 0.292 $\mu$m $\times$ 0.268 $\mu$m (for $\lambda=780$ nm), and 0.495 $\mu$m $\times$ 0.45 $\mu$m  (for $\lambda=1310$ nm). (d) Net transversal chiral and achiral forces that a quasi-CP mode in a strip waveguide exert on a particle depending on the particle size. (e) Cross-section of the strip waveguide showing the position where the forces are evaluated for (c) and (d): at a vertical distance over the top of the waveguide equal to the radius of the particle, and at a horizontal distance equal to the fourth of the waveguide width from the center of the waveguide.}
    \label{fig:cross-section}
\end{figure}

We consider two different approaches to sorting chiral nanoparticles throughout the transversal plane ($xy-$plane) around the waveguide: the fundamental quasi-transverse electric mode (quasi-TE mode) for horizontal sorting and a quasi-circularly polarized mode (quasi-CP mode) for attractive-repulsive sorting. The electric field intensity and polarization of the quasi-TE mode and the quasi-CP mode are plotted in Fig.~\ref{fig:cross-section}(b). A more detailed decomposition of the electric and magnetic fields for the quasi-TE, quasi-TM, and quasi-CP modes can be found in the Appendix~\ref{sec:modes_EH} (Fig.~\ref{fig:Strip_mode_fields}).

\subsection{Quasi-TE mode}
The fundamental quasi-TE mode (or TE$_0$) in a strip waveguide is characterized by the transverse electric field mainly pointing along the horizontal plane ($x-$axis) and the transverse magnetic field mainly polarized along the vertical plane (along $y-$axis). This guided mode exhibits a non-zero longitudinal component of the electric and magnetic field as a consequence of confining the wave inside a waveguide. The longitudinal component oscillates out-of-phase in comparison with the transversal components. This results in a transverse spin \cite{Espinosa-Soria2016}, which can be interpreted as the quantum spin Hall effect of light \cite{Bliokh2015} and gives rise to transversal chiral forces.

\begin{figure}
    \centering
    \includegraphics[width=0.89\textwidth]{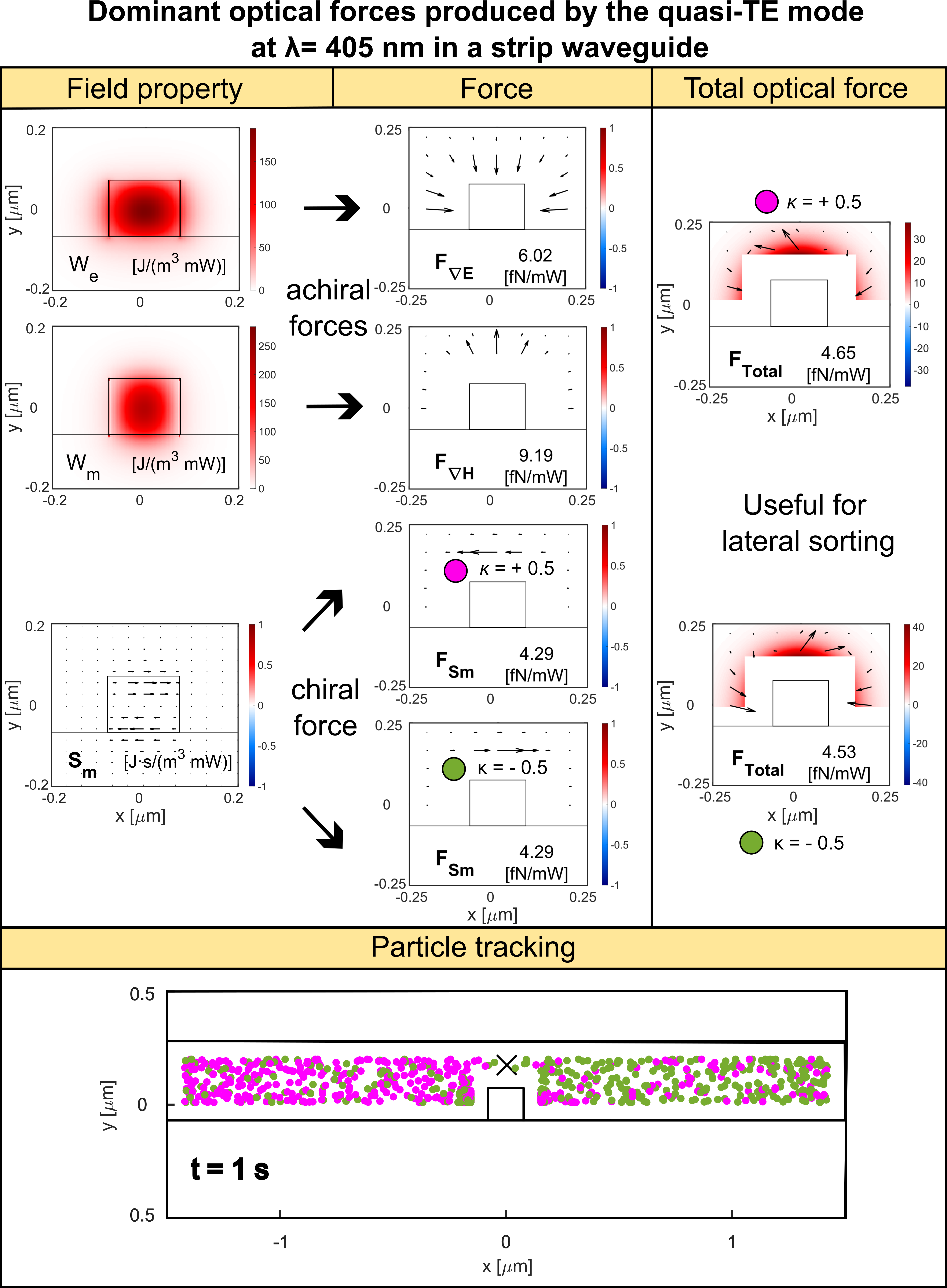}
    \caption{Field properties ($W_e$, $W_m$ and $\vc{S}_m$) that generate the dominant forces of the quasi-TE mode in a strip waveguide (0.151 $\mu$m wide $\times$ 0.139 $\mu$m thick) at $\lambda=405$ nm. The force stemming from the $\vc{S}_m$ as well as the total optical force are represented for both chiralities of the particle ($\kappa = \pm\ 0.5$). All forces are calculated for a particle of 80 nm radius. The axes of all graphs refer to the $x-$ and $y-$coordinate measured in $\mu$m units. The arrow map represents the transversal components of the vectorial quantities being plotted, and the colormap represents the scalar quantity or the $z-$component of the vectorial quantity being plotted.  The particle tracking graph shows the last position of 500 particles per enantiomer inside the microchannel (3 $\mu$m wide $\times$ 0.5 $\mu$m thick) after 1 s of motion given the total optical force field produced by the waveguide. The initial position of all particles is marked by the $\times$ symbol.}
    \label{fig:TE_405}
\end{figure}

The optical forces that a quasi-TE mode exerts on a particle are computed for a wavelength of 405 nm, and a particle radius of 80 nm. The field properties that are responsible for the dominant transversal forces in this system are the electric energy density $W_e$, the magnetic energy density $W_m$, and the magnetic spin $\vc{S}_m$. These field properties together with their respective force are plotted in Fig.~\ref{fig:TE_405}. The electric gradient force attracts any particle toward the sidewalls of the waveguide, whereas the magnetic gradient force repels any particle from the top of the waveguide. The magnetic spin force moves the $(+)$-particles towards the left and the $(-)$-particles towards the right. Over the top of the waveguide, the achiral gradient electric and gradient magnetic forces have opposite directions, thus reducing the strength of the total achiral force. That is why the chiral force magnitude ($\sim4.29$ fN/mW) is of the same order as the total achiral force ($\sim9.19-6.02=3.17$ fN/mW), resulting in the total force along the $x-$axis changing sign for opposite enantiomers over the top of the waveguide, thus, pushing enantiomers to opposite sides along the $x-$direction.

The motion of 500 particles per enantiomer is tracked individually for 1 s given the total optical force field shown in Fig.~\ref{fig:TE_405}. The microchannel dimensions (width $\times$ height) are 3 $\mu$m $\times$ 0.35 $\mu$m. The resulting final positions of the particles are plotted in Fig.~\ref{fig:TE_405} for both enantiomers. From among the 500 $(+)$-particles 64\% end up on the left side ($x<0$) and 59\% of $(-)$-particles end up on the right side ($x>0$), yielding an enantiomer fraction (EF) of 61.0\% and 62.1\% respectively. The (+)-EF is calculated at $x<0$ and the ($-$)-EF is calculated at $x>0$. Particle tracking simulations assuming powers stronger than 20 mW, and therefore stronger forces, suggest that the EF increase with power injected into the mode yielding 76\% for 50 mW and 85\% for 100 mW. 

A way to increase the strength of chiral forces over the achiral forces is by exploiting the wavelength dependency of the forces. Some of the forces depend on the ratio $r/\lambda$, as previously discussed for Eq.~(\ref{eq:all_forces}). This dependency is studied in Fig.~\ref{fig:cross-section}(c) for the quasi-TE mode, where the total chiral and achiral forces along $x-$ direction, which is the sorting direction, are represented with respect to the particle radius for three wavelengths (405 nm, 780 nm, and 1310 nm). In all the cases, the forces are evaluated at a point positioned at a vertical distance equal to the particle's radius over the top of the waveguide, and at a horizontal distance equal to 1/4 of the waveguide’s width from the center; as shown in the inset in Fig.~\ref{fig:cross-section}(e). That vertical distance is the minimum distance the particle can be at due to its size. Fig.~\ref{fig:cross-section}(c) shows that there is a particle size range over which chiral forces become larger than achiral forces along the $x-$direction: 75$-$138 nm for $\lambda=405$ nm, 148$-$288 nm for $\lambda=780$ nm, and 235$-$460 nm for $\lambda=1310$ nm. In fact, there is a specific radius that maximizes the ratio of chiral force over achiral force within those intervals. In these ranges, the spin magnetic force is stronger than the other achiral forces along the $x-$direction. However, for smaller radii, the achiral magnetic gradient becomes dominant, and for larger radii, the sum of the achiral flow force due to the imaginary part of the Poynting vector and the achiral magnetic gradient becomes dominant instead. In addition, these intervals suggest that operating at shorter wavelengths is more suitable for sorting smaller chiral particles.
 
\begin{figure}
    \centering
    \includegraphics[width=0.85\textwidth]{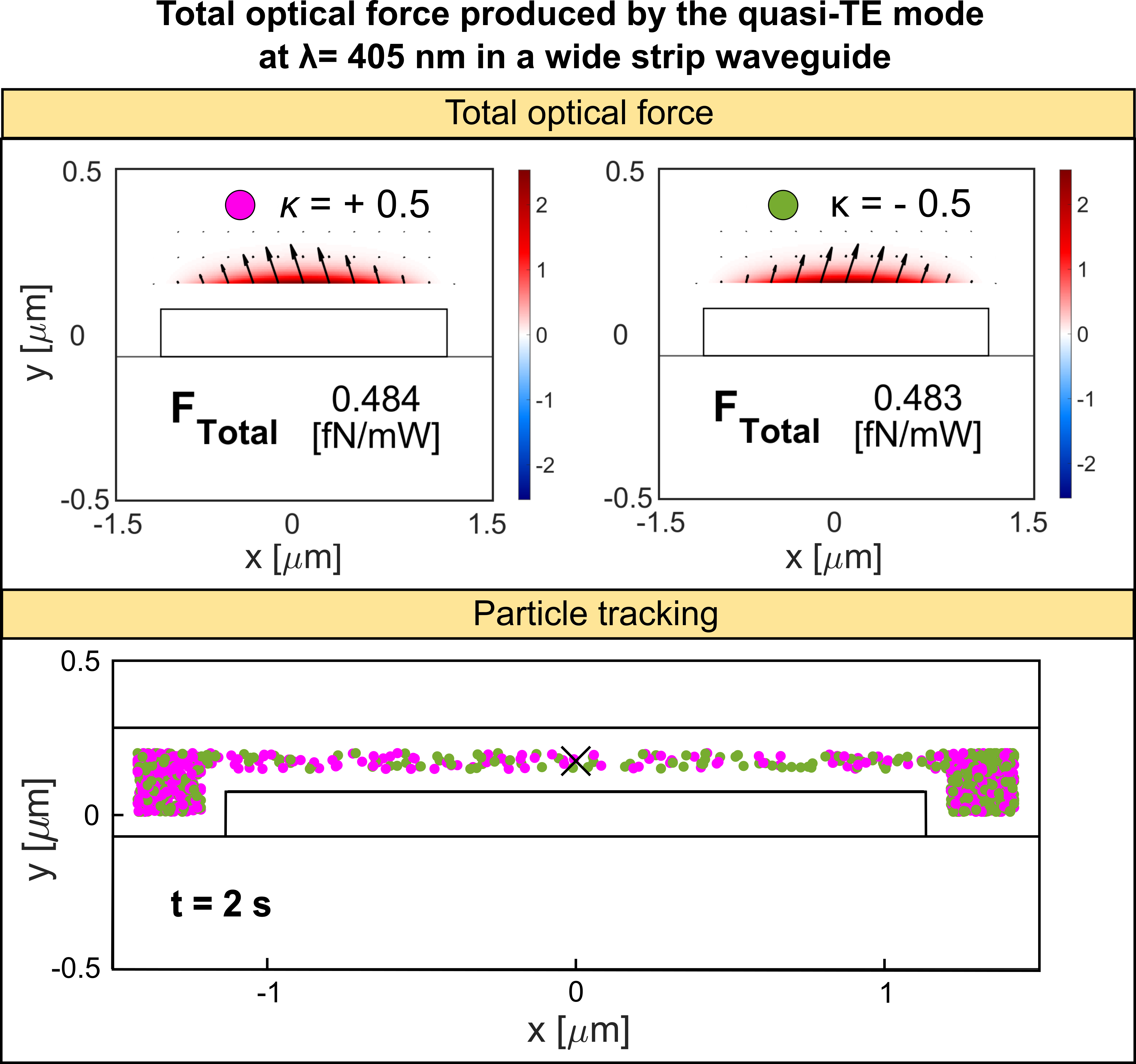}
    \caption{The total optical force (transversal in arrow map, and longitudinal in colormap) that a quasi-TE mode in a wide strip waveguide (2.270 $\mu$m wide $\times$ 0.139 $\mu$m thick) at $\lambda=405$ nm exerts on an 80 nm-radius particle is shown for both chiralities of particle $\kappa \pm 0.5$. The particle tracking graph shows the last position of 500 particles per enantiomer inside the microchannel (3 $\mu$m wide $\times$ 0.5 $\mu$m thick) after 2 s of motion given the total optical force field shown in the same figure. The initial position of all particles is marked by the $\times$ symbol.}
    \label{fig:TE_wide_405}
\end{figure}

Another way to facilitate the horizontal chiral separation is to lower the strength of the achiral gradient forces along the $x-$axis. This can be achieved by making the waveguide wider. The power is thus spread over a larger area, thereby reducing the gradient of the fields and their respective achiral gradient forces along the $x-$axis. In addition, the smallest particle size for which chiral and achiral forces have the same magnitude reduces down to 40 nm for $\lambda=405$ nm. This comes at the expense of reducing the strength of the chiral optical force too ($\sim0.484$ fN/mW), as shown in Fig.~\ref{fig:TE_wide_405} for a waveguide 2.27 $\mu$m wide $\times$ 0.5 $\mu$m thick, since the power of the mode has been distributed over a larger cross-section. The particle tracking simulation inside a microchannel of the same dimensions as before (3 $\mu$m wide $\times$ 0.5 $\mu$m thick), yielded practically the same value of EF ($\sim61$\%/$\sim62$\%) for (+)/($-$)-enantiomers as with the narrower waveguide despite the magnitude of the forces being one order of magnitude less. This might be because the wider waveguide allows the interaction between the particle and optical force for a longer time because the waveguide width is larger. In spite of yielding the same enantiomeric separation capability, this wider configuration should be easier to implement experimentally due to the larger area of interaction between the mode and the particles.



\subsection{Quasi-CP mode}
We refer to the quasi-CP mode as a guided mode that is obtained by the superposition of the TE$_0$ mode and the TM$_0$ mode delayed by a phase shift of $90^\circ$. As the electric field is predominantly horizontally polarized in the TE$_0$ mode and predominantly vertically polarized in the TM$_0$ mode, the combination originates a guided mode with an effective circular polarization and, therefore, local helicity \cite{Vázquez‐Lozano2020}. The waveguide width and thickness are chosen so that the TE$_0$ mode and the TM$_0$ mode are degenerate, i.e. both modes exhibit the same effective refractive index ($\Delta n = n_{\text{TE}}-n_{\text{TM}} = 0$) at the target wavelength. This degeneracy allows the circular polarization of the mode to be maintained along the waveguide. 

\begin{figure}
    \centering
    \includegraphics[width=0.89\textwidth]{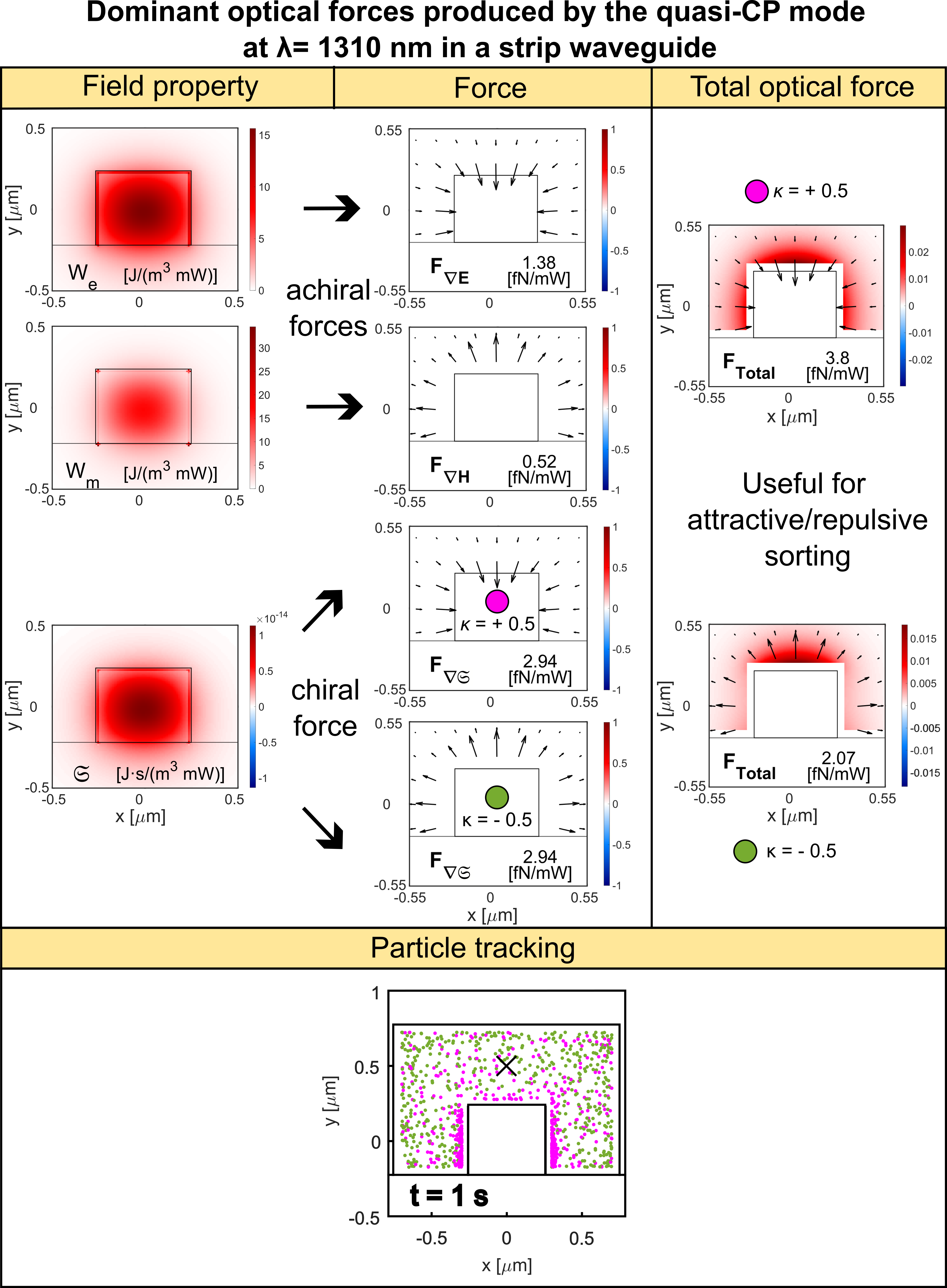}
    \caption{Field properties ($W_e, W_m$ and $\mathfrak{G}$) that generate the dominant forces of the quasi-CP mode in a strip waveguide (0.495 $\mu$m wide $\times$ 0.450 $\mu$m thick). The force stemming from the gradient of $\mathfrak{G}$ as well as the total force are represented for both chiralities of the particle ($\kappa=\pm 0.5$). All forces are calculated for a particle of 52 nm radius. The arrow map represents the transversal components of the vectorial quantities being plotted, and the colormap represents the scalar quantity or the z-component of the vectorial quantity being plotted. The particle tracking graph shows the last position of 500 particles per enantiomer inside the microchannel (1.5 $\mu$m wide $\times$ 1 $\mu$m thick) after 1 s of motion given the total optical force field produced by the waveguide. The initial position of all particles is marked by the $\times$ symbol.}
    \label{fig:CP_1310}
\end{figure}

We computed the dominant achiral and chiral forces of this system for a particle of 52 nm radius and a wavelength of 1310 nm (see Fig.~\ref{fig:CP_1310}). The field of the quasi-CP mode exhibits an intrinsic local helicity $\mathfrak{G}$, which naturally diminishes along the evanescent field beyond the waveguide core. This is a robust way of achieving a helicity gradient, which is in fact the predominant chiral force in this system. This force is particularly interesting because it depends only on the chiral polarizability, which means that it grows with the radius of the particle as $r^{3}$ instead of $r^{6}$, and does not depend explicitly on the wavelength. The total optical force exhibits opposite directions for opposite chiralities, being attractive towards the waveguide for the (+)-particles (with value $\sim3.8$ fN/mW) and repulsive for the $(-)$-particles (with value $\sim2.07$ fN/mW).

In order to identify the possible range of sizes that might be easier to sort with the quasi-CP compound mode, we performed the study of the chiral forces dependence on the particle’s size at the wavelength of 1310 nm. As shown in Fig.~\ref{fig:cross-section}(d), the net transversal chiral force (along $x-$ and $y-$axis) dominates over the net transversal achiral force, up to a maximum size (409 nm) where the combination of the forces due to the real and imaginary parts of the Poynting vector and the rotational of the spin overtake the helicity gradient force. Crucially, this means that the size range has no minimum radii where the chiral force is smaller than the achiral force, as it occurs for the quasi-TE mode. This means that we can expect to reach enantioseparation for smaller particles when compared to the TE mode, even down to radii around 1 nm (molecular size), making this force the most promising for sorting molecules. The particle tracking results inside a microchannel (1.5 $\mu$m wide $\times$ 1 $\mu$m thick) for the system in Fig~\ref{fig:CP_1310} show that 61.4\% of the 500 (+)-particles become `attached' to the waveguide within a radius of 425 nm measured from the center of the waveguide, and that 90.2\% of the 500 $(-)$-particles are repelled. The (+)-EF within a radius of 425 nm is 86.4\% and the ($-$)-EF outside the radius of 425 nm is 70.0\%.

 
\section{Conclusions}
In conclusion, we have identified an opportunity for transversal enantioseparation via optical forces in photonic integrated waveguides utilizing distinct mechanisms depending on particle size and operating wavelength. To this end, we have used the most straightforward waveguide structure: a strip SiN waveguide placed on a silica substrate. At short wavelengths such as around 405 nm, the spin magnetic force arising from a quasi-TE mode is in principle strong enough to sort particles larger than 80 nm radius under in less than 1 s. At longer wavelengths, such as 1310 nm, the helicity gradient force stemming from a quasi-CP mode can sort particles of size down to 52 nm radius under time spans bekow 1 s. In contrast to other approaches using optical waveguides \cite{Fang2021,Liu2023,Fang2023}, our waveguides are longitudinally invariant, meaning that the optical chiral forces could be exerted over long distances (cm-scale), thus facilitating practical enantioseparation with realistic optical powers in the chip (20 mW). These findings underscore the potential of optical forces generated in integrated waveguides in facilitating enantioseparation within the specified parameters of particle size and wavelength.


\section*{Acknowledgements}
The authors acknowledge financial support from the European Commission under contract EIC Pathfinder CHIRALFORCE 101046961. A. M. acknowledges partial funding from the Generalitat Valenciana under the NIRVANA grant (PROMETEO Program, CIPROM/2022/14) and F. J. R-F. acknowledges partial funding from Innovate UK Horizon Europe Guarantee (UKRI project 10045438).

\section{APPENDIX}
\subsection{Particle tracking algorithm} \label{sec:particle_tracking}
As shown in the main text, the motion of a Brownian particle immersed in a fluid and subjected to a force field $\vc{F}=(F_x,F_y,F_z)$ is governed by the overdamped Langevin equation. Its solution can be given by the Euler-Maruyama integration scheme \cite{Kloeden1992,Kieninger2021} to track the location of the particle in time:
\begin{equation} \label{eq:maruyama_SI}
\begin{split}
    x^{(m+1)} & =  x^{(m)} + \frac{F_x^{(m)}}{\gamma}\Delta t + \sqrt{2D_{x,\perp}^{(m)}\Delta t} \; N(0,1) \\
    y^{(m+1)} & = y^{(m)} + \frac{F_y^{(m)}}{\gamma}\Delta t + \sqrt{2D_{y,\perp}^{(m)}\Delta t} \; N(0,1) \\
    z^{(m+1)} & = z^{(m)} + \frac{F_z^{(m)}}{\gamma}\Delta t + \sqrt{2D_{z,||}^{(m)}\Delta t} \; N(0,1)
\end{split}
\end{equation}
\noindent where $N(0,1)$ represents a standard normal distribution with $0-$mean and $1-$variance, and the force field $\vc{F}$ is evaluated at the location of the particle ($x^{(m)},y^{(m)},z^{(m)}$) at each instant of time. The time step used in all the particle tracking simulations is $\Delta t = 10$ $\mu s$.

The diffusion coefficients are modified, from the bulk value $D$, to account for the hydrodynamic interaction between the particle and the enclosing boundaries. The modifications are different depending on whether the particle moves along the directions parallel ($||$) or perpendicular ($\perp$) to a non-slip planar boundary \cite{Swan2007}:
\begin{equation}
\begin{split}
    D_{\perp}(h) & = D\left[1 - \frac{9}{8}\left(\frac{r}{h}\right) + \frac{1}{2}\left(\frac{r}{h}\right)^3 -\frac{1}{8}\left(\frac{r}{h}\right)^5  \right] \\
    D_{||}(h) & = D\left[1 - \frac{9}{16}\left(\frac{r}{h}\right) + \frac{1}{8}\left(\frac{r}{h}\right)^3 -\frac{1}{16}\left(\frac{r}{h}\right)^5  \right]
\end{split}
\end{equation}
\noindent where $h$ is the distance from the center of the particle to the wall, and $r$ is the radius of the spherical particle. We used the $\perp-$correction for the $x-$ and $y-$directions and the $||-$correction for the $z-$direction. We do not allow the particles to come closer than a radius-distance from the wall, i.e. $h\geq r$. This bounds the values of $r/h$ between 0 and 1, and therefore: $D_{\perp}\in [D/4,D]$, $D_{||}\in [D/2,D]$. These modified coefficients are obtained for the case of a particle moving nearby a single planar boundary. However, in our waveguide system we have multiple boundaries, defined by the ceiling and sidewalls of the microchannel as well as top and sidewalls of the waveguide core and top of the substrate. We assume the modified expressions are still valid. We use the $D_{\perp}$ expression for calculating the diffusion coefficient along $x-$ and $y-$direction, and use the $D_{||}$ expression along $z-$direction. We take $h$ as the distance to the nearest boundary in the direction of motion.  For instance, for the calculation of $D_{y,\perp}$ we compute $h$ by measuring the distance from the particle position to the ceiling of the microchannel and compare it to the distance to the substrate or top of the waveguide core (depending on whether the particle's position is above the substrate or above the waveguide core). Whichever distance is shorter, that is the value of $h$.

The force field used in the algorithm is obtained for a system that is not enclosed by a microchannel. However, we assume the presence of the microchannel does not modify the guided modes, and thus, the force field produced by the waveguide throughout the surrounding medium (water, $n=1.33$). This approximation is valid due to the higher refractive index of the of the waveguide core (SiN, $n\sim$2) with respect to that of the microchannel material (SiO$_2$, $n\sim$1.45), which ensures the guidance of the mode along the core, and because the microchannel walls (boundaries) are separated by a distance $\geq \lambda/1.33$ from the waveguide walls.

Once the particle tracking simulation is finished, we do statistical analysis with the last positions of the 500 particles for both enantiomers, to compute the enantiomer fraction. Given the number of (+)-particles, $N_+$, and the number of ($-$)-particles, $N_-$, inside a region of space, we define the enantiomer fraction (EF) as \cite{Smith2009}:
\begin{equation}
\begin{split}
    (+)\text{-EF} &= \frac{N_+}{N_+ + N_-} \\
    (-)\text{-EF} &= \frac{N_-}{N_+ + N_-}
\end{split}
\end{equation}

Since the chiral forces separate opposite enantiomers in opposite directions the (+)-EF and the ($-$)-EF are calculated in different regions of space. For instance, for the quasi-TE mode the (+)-EF is calculated for the region $x<0$ whereas the ($-$)-EF is obtained for the region $x>0$. And, for the quasi-CP mode the (+)-EF is calculated for the region $x^2+y^2<R^2$ whereas the ($-$)-EF is obtained for $x^2+y^2>R^2$, where $R$ is an arbitrary radius that defines a circular region from the center of the waveguide.

\subsection{Sorting time for a cloud of enantiomers in a fluid}\label{sec:sorting_time}
%
%
We derive the expression for calculating the time needed for a chiral optical force, $F$, to separate two clouds of opposite enantiomers a distance $\Delta x$. This has been previously done by Kravets \textit{et al.} \cite{Kravets2019}, but we include it here for completeness. The chiral force (uniform in space and constant in time) moves each enantiomer cloud into opposite directions a distance $d_{\text{opt}}$ from their initial position
\begin{equation}
    d_{\text{opt}}=\frac{F}{\gamma}t
\end{equation}
\noindent where $\gamma$ is the friction coefficient of the particle motion in the fluid. Let us assume the chiral force moves the $(+)$-enantiomer cloud towards the left ($-x$ direction) and the $(-)$-enantiomer cloud towards the right ($+x$ direction), following Fig.~\ref{fig:sorting}(a) schematic. Therefore, the separation between the center of mass of both clouds has increased $2d_{\text{opt}}$. In addition to this displacement, the radius for each enantiomer cloud is expected to increase on average an extra distance $d_\text{B}$ from its initial extension $L_0 /2$ due to Brownian motion
\begin{equation}
    d_\text{B}= \sqrt{2Dt}
\end{equation}
\noindent where $D=k_\text{B}T/\gamma$ is the diffusion coefficient, $k_\text{B}$ is the Boltzmann constant and $T$ is the absolute temperature of the fluid.
These two simultaneous processes of motion are schematized in Fig.~\ref{fig:sorting}, where by comparing the defined lengths one can see that the condition to achieve a separation of $\Delta x$ in a sorting time $t$ is given by:
\begin{equation}
    2d_{\text{opt}} = L_0 + 2d_\text{B} + \Delta x
\end{equation}
\begin{equation}
       2\frac{F}{\gamma}t = L_0 +  2\sqrt{2D}\sqrt{t} +\Delta x 
\end{equation}

This equation can be solved as a second-degree equation on the variable $\sqrt{t}$, whose solution is
\begin{equation}
    \sqrt{t_{\text{sort}}}= \frac{\gamma}{F}\sqrt{\frac{D}{2}}\left(1 +  \sqrt{1 + \frac{F(L_0+\Delta x)}{D\gamma}} \right)
\end{equation}
\noindent where we have taken the positive square root solution for avoiding unphysical negative time. By squaring this last expression and substituting $D=k_B T/\gamma$ and $\gamma=6\pi\eta r$, the sorting time can be obtained:
\begin{equation}
t_{\text{sort}}= 3 \pi \eta k_\text{B}T\frac{r }{F^2}\left( 1 + \sqrt{1 + \frac{F(L_0+\Delta x)}{k_\text{B}T}} \right) ^2
\end{equation}

\subsection{Electric and Magnetic field of the guided modes}
\label{sec:modes_EH}
\begin{figure}[!htb]
    \centering
    \includegraphics[width=0.9\textwidth]{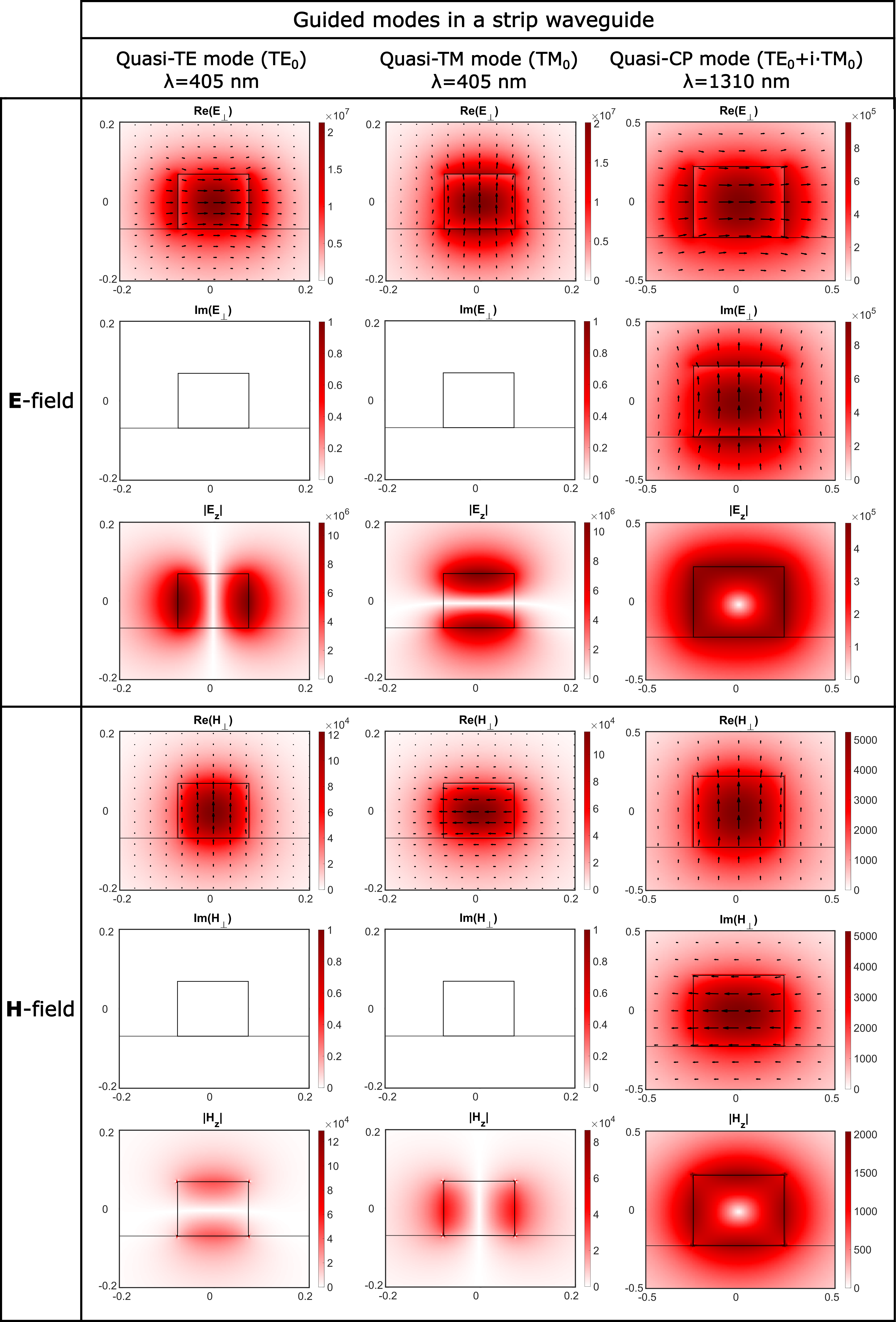}
    \caption{Electric and Magnetic field patterns of the guided modes discussed in the main text for a strip waveguide. The fields are split into the transversal ($\perp$) and longitudinal ($z$) components. The transversal components of the TE$_0$ and TM$_0$ modes lack imaginary part, whereas the quasi-CP mode does due to the circular polarization of the mode induced by the phase delay of 90$^\circ$.}
    \label{fig:Strip_mode_fields}
\end{figure}



\bibliography{sn-bibliography}
\end{document}